# Design and Implementation of BNN-Based Object Detection on FPGA

Xuyu Zhao[1], Yunpeng Wu[2], Mengyuan Zhu[3], Haoyu Huang[2], Xiaoyu Xu[4], Yanjing Li[2], Gaolong Zhang[1,5], Baochang Zhang[2]

[1] School of Physics, Beihang University, Beijing, China
[2] School of Artificial Intelligence, Beihang University, Beijing, China
[3] China Information Technology Designing and Consulting Institute Company Ltd, Beijing , China
[4] State Key Laboratory of Radiation Medicine and Protection, School of Radiation Medicine and Protection, Collaborative Innovation Center of Radiological Medicine of Jiangsu Higher Education Institutions, Soochow University, Suzhou, China
[5] State Key Laboratory of Complex & Critical Software Environment (SKLCCSE) Beihang University, Beijing, China

Abstract. This paper implements a Binary Neural Network (BNN) based YOLOv3-tiny-like object detector on a low-cost FPGA. The network takes 320×320×3 RGB images as input. Its main convolution layers use 1-bit weights and 8-bit activations, while Conv1 and the final detection head use fixed-point standard convolutions. From the trained ONNX model, weights, biases, and quantization parameters are extracted, converted to fixed point, packed into COE files, and stored in Vivado BRAM ROMs. The hardware is written fully in Verilog RTL and includes padding, line buffering, binary convolution, quantization post-processing, max pooling, and detection-head computation. For layers where Mul_prev is indexed by input channel and Div_current by output channel, Mul_prev is fused into the BNN PE so that channel-wise compensation is applied during accumulation. On VOC, the model obtains 39.6% mAP50 with 0.098 GFLOPs and 0.74 M parameters. RTL simulation shows that the final raw detection output reaches a correlation coefficient of 0.999964 and a mean absolute error of 0.020027 against the corresponding ONNX node.

## 1 Introduction

Convolutional neural networks are widely used in object detection and edge vision[1-2]. Because detection networks must preserve high-resolution feature maps, their direct deployment on low-cost embedded platforms is bounded by on-chip memory, compute, and bandwidth — motivating low-bit quantization at the hardware level.

FPGAs provide customizable datapaths and fine-grained parallelism, making them well suited to bit-width-specific accelerators that ASIC and GPU pipelines cannot match. For resource-limited boards such as PYNQ-Z2, deployment must control weight storage, activation buffering, DSP usage, and control complexity. Low-bit quantization is therefore a practical way to reduce hardware cost.

Fully binary networks greatly reduce storage and multiplication cost, but their feature representation can be unstable for detection. W1A8 quantization uses 1-bit weights and 8-bit activations. It replaces weight multiplication with signed addition or subtraction while keeping an activation range of 0 to 255, which gives a more suitable accuracy-cost trade-off for object detection deployment on resource-limited FPGAs.

This work develops the mapping from a trained ONNX model to FPGA RTL. It covers parameter parsing, fixed-point conversion, COE generation, BRAM ROM deployment, streaming convolution, and layer-wise verification. Handwritten Verilog RTL is used instead of HLS to control bit widths, ROM latency, pipeline states, and channel-wise scaling.

This paper presents a complete deployment structure of a YOLOv3-tiny-like W1A8 detector on XC7Z020, integrates input-channel scaling compensation into the W1A8 PE, and verifies the RTL implementation against ONNX Runtime through layer-wise comparison, model-level metrics, resource use, and detection results.

## 2 Related Work

### 2.1 Object Detection and Lightweight YOLO

The YOLO family treats detection as one-stage regression and provides a useful balance between speed and accuracy[3-5]. YOLOv3 introduces multi-scale prediction, and YOLOv3-tiny reduces layers and channels for edge scenarios. This work adopts a YOLOv3-tiny-like structure, further reduces channel count and weight precision, and targets the limited on-chip resources of XC7Z020.

### 2.2 Low-Bit Quantization and Binary Neural Networks

BinaryConnect, BNN, XNOR-Net, and DoReFa-Net show that low-bit weights and activations can compress neural networks effectively[6-9]. ReActNet, LSQ, and integer-only inference studies further highlight the roles of activation behavior, learnable steps, and fixed-point scaling[10-12]. This work uses W1A8 to keep the storage benefit of 1-bit weights while reducing the accuracy loss of W1A1 detection.

### 2.3 FPGA Acceleration of Neural Networks

FPGA neural network accelerators usually rely on data reuse, on-chip buffering, parallel computation, and pipeline scheduling[13]. Automated tool chains such as FINN, FINN-R, DNNWeaver, and hls4ml support binary, quantized, or high-level deployment[14-17]. We instead use handwritten RTL, which exposes ROM-read latency, per-stage bit widths, and per-channel scaling fusion — fine-grained controls needed to fit the 4.9 Mb BRAM budget of XC7Z020.

### 2.4 BNN and YOLO Deployment on FPGA

Prior FPGA studies implement binary convolution accelerators and YOLO-oriented accelerators[18-22]. This paper focuses on the parameter organization, fixed-point data path, and RTL-ONNX numerical alignment of a W1A8 detector on XC7Z020.

## 3 W1A8 Quantized Model Structure

### 3.1 Input, Output, and Layer Structure

The deployed model is a trained ONNX detector. Software inference normalizes RGB pixel values from [0, 255] to [0, 1]. RTL approximates the input as Q0.8 fixed-point values so that the hardware scale matches the software input. The model input is 320×320×3, and the final output is a 10×10×75 tensor in y/x/channel order.

The model contains eleven main convolution layers from Conv1 to Conv11. Conv2 to Conv10 use W1A8 quantization. Conv1 and Conv11 use fixed-point standard convolution to reduce input-side and output-side errors. MaxPool is used only after Conv1, Conv2, Conv3, Conv4, and Conv7. Table 1 gives the layer configuration.

Table 1. Structure and quantization configuration of the main convolution layers

| Layer | Convolution type | Channels | Kernel size | Following operation | Fixed-point/quantization format |
|---|---|---|---|---|---|
| Conv1 | Standard convolution | 3→16 | 3×3 | Post + MaxPool | W: Q5.11; B: Q2.14 |
| Conv2 | W1A8 | 16→32 | 3×3 | Post + MaxPool | 1-bit W; 8-bit A |
| Conv3 | W1A8 | 32→64 | 3×3 | Post + MaxPool | 1-bit W; 8-bit A |
| Conv4 | W1A8 | 64→128 | 3×3 | Post + MaxPool | 1-bit W; 8-bit A |
| Conv5 | W1A8 | 128→128 | 3×3 | Post | 1-bit W; 8-bit A |
| Conv6 | W1A8 | 128→128 | 3×3 | Post | 1-bit W; 8-bit A |
| Conv7 | W1A8 | 128→128 | 3×3 | Post + MaxPool | 1-bit W; 8-bit A |
| Conv8 | W1A8 | 128→128 | 3×3 | Post | 1-bit W; 8-bit A |
| Conv9 | W1A8 | 128→64 | 1×1 | Post | 1-bit W; 8-bit A |
| Conv10 | W1A8 | 64→64 | 3×3 | Post | 1-bit W; 8-bit A |
| Conv11 | Standard convolution | 64→75 | 1×1 | Output | W: Q1.15; B: Q4.12 Output: signed Q*.15 |

Table 1 shows that early layers have larger feature maps and higher buffer pressure, while later layers reduce the feature-map size to 10×10. Weights, biases, and quantization parameters are converted to COE files and stored in BRAM ROMs. Line buffers and outputs are organized by window size, channel count, and parallelism.

Table 2 estimates the storage pressure of each layer. The estimates are based on feature-map size, channel count, data precision, and kernel size. Actual BRAM usage also depends on ROM packing, memory-depth alignment, port organization, and control logic.

Table 2. Estimated deployment storage cost of each layer

| Layer | Type | Input→Output | Kernel size | H×W | Estimated output/line buffer | Weight storage |
|---|---|---|---|---|---|---|
| L0 | Standard convolution | [3,320,320] → [16,320,320] | 3×3 | 320×320 | 2×320×16 = 10.0KB | 2.0KB |
| L1 | MaxPool | [16,320,320] → [16,160,160] | 2×2 | 160×160 | 3×160×16 = 7.5KB | 0 |
| L2 | W1A8 Conv | [16,160,160] → [32,160,160] | 3×3 | 160×160 | 2×160×32 = 10.0KB | 1.3KB |
| L3 | MaxPool | [32,160,160] → [32,80,80] | 2×2 | 80×80 | 3×80×32 = 7.5KB | 0 |
| L4 | W1A8 Conv | [32,80,80] → [64,80,80] | 3×3 | 80×80 | 2×80×64 = 10.0KB | 3.8KB |
| L5 | MaxPool | [64,80,80] → [64,40,40] | 2×2 | 40×40 | 3×40×64 = 7.5KB | 0 |
| L6 | W1A8 Conv | [64,40,40] → [128,40,40] | 3×3 | 40×40 | 2×40×128 = 10.0KB | 12.0KB |
| L7 | MaxPool | [128,40,40] → [128,20,20] | 2×2 | 20×20 | 3×20×128 = 7.5KB | 0 |
| L8-L10 | W1A8 Conv | [128,20,20] → [128/64,20/10,20/10] | 3×3/1×1 | 20×20/10×10 | 5.0-7.5KB | 2.5-21.0KB /layer |
| L11 - L15 | Max-Pool/Conv/Detect | [128,10,10] → [75,10,10] | 2×2/3×3/1×1 | 10×10 | 640B-29.3KB | 0-21.0KB /layer |

Table 2 shows that early layers mainly consume buffers, whereas later layers increase weight-ROM pressure. W1A8 quantization reduces main-layer weight storage, and the streaming data path reduces the need to store complete intermediate feature maps.

### 3.2 W1A8 Quantization and Fixed-Point Inference Mapping

The main convolution layers use W1A8 quantization to reduce storage and computation. The 1-bit weights reduce ROM capacity and bandwidth, while 8-bit activations preserve a useful dynamic range for detection features. Conv1 and Conv11 remain fixed-point standard convolutions to reduce errors at the input and output sides.

For weight quantization, the weights in the main convolution layers are binarized into {−1,+1}. Let the real-valued weight be $w$ and the binary weight be $w_b$. The binarization is written as

$$w_b = sign(w), w_b \in -1, +1 \qquad (3-1)$$

Training keeps real-valued weights and uses binary weights in the forward pass. The straight-through estimator approximates the gradient of the sign function. During inference, only binary weights are stored, reducing the theoretical weight storage of the main convolution layers to 1/32 of 32-bit precision. Because the design is W1A8 rather than W1A1, convolution uses sign-controlled accumulation instead of XNOR-popcount.

For W1A8 convolution, let $a_i$ be an 8-bit non-negative activation and $s_{o,i}$ be the binary weight sign corresponding to output channel o. The convolution accumulation is expressed as

$$y_o = \sum_i s_{o,i} a_i, s_{o,i} \in -1, +1, a_i \in [0,255] \qquad (3-2)$$

Index $i$ covers both input channels and kernel positions. W1A8 convolution therefore replaces general multiplication with sign-controlled addition and subtraction. This form maps to LUTs, adders, and registers and helps reduce DSP usage.

For activation quantization, this work adopts learned step-size quantization to map ReLU activations to 8-bit unsigned integers. Because ReLU outputs are non-negative, the activation range is set to [0,255]. Let the floating-point activation be $x$ and the quantization step size be $s_a$. The activation quantization is written as

$$q_a = \text{clip}\left(\text{round}\left(\frac{x}{s_a}\right), 0{,}255\right) \qquad (3-3)$$

Here, $q_a$ is the 8-bit activation, round denotes rounding, and clip denotes clipping. During training, the step size $s_a$ is optimized with the network parameters. During inference, the learned step size becomes fixed-point scaling, rounding, and clipping in RTL.

Floating-point scales in the inference graph are converted to fixed-point operations. For W1A8 layers, accumulation results pass through scaling, bias correction, rounding, and clipping to generate the next 8-bit activation. The final detection head keeps the raw tensor for comparison with ONNX.

The hardware does not merge all scales into one constant. In several layers, Mul_prev is indexed by input channel, while Div_current is indexed by output channel. A single scale would ignore input-channel contribution differences and introduce additional error.

To avoid this problem, Mul_prev is integrated into the W1A8 PE so that input-channel compensation is completed during accumulation. Let $m_i$ be the previous scale corresponding to the i-th input channel. The compensated accumulation is written as

$$y_o = \sum_i s_{o,i}(m_i a_i) \qquad (3-4)$$

In RTL, $m_i$ is fixed-point and participates in PE accumulation, preserving input-channel scaling differences. Div_current is handled in post-processing for output-channel scaling, rounding, and clipping. This split improves consistency with ONNX without greatly increasing post-processing complexity.

# 4 Parameter Extraction and Fixed-Point Deployment

The extraction script reads weights, biases, Div tensors, Mul tensors, and related parameters from the ONNX model. It rearranges them according to the channel order and ROM access order used by RTL, then performs fixed-point conversion and COE packing.

Conv1 uses Q5.11 weights and Q2.14 biases. Conv2 to Conv10 use 1-bit weights and 8-bit activations. Conv11 uses Q1.15 weights and Q4.12 biases and outputs 32-bit signed fixed-point values with 15 fractional bits, recovered as signed_int32 / 2^15. The formats are chosen according to parameter ranges and output ranges, and the packed parameters follow the RTL ROM address order.

All weights, biases, and quantization parameters are stored in BRAM ROMs. Because the Vivado BRAM ROM core does not enable Primitives Output Register, the RTL state machine schedules address issue, data latch, and computation start according to the actual ROM read latency.

Table 3. Parameter extraction and COE generation configuration

| Parameter type | Source | Fixed-point/processing method | Hardware use |
|---|---|---|---|
| Conv1 weights and biases | ONNX initializer | Q5.11 / Q2.14 | First 3×3 standard convolution |
| W1A8 weights | ONNX initializer | 1-bit sign packing | Sign-controlled add/sub accumulation from Conv2 to Conv10 |
| Div_current | ONNX quantization-related tensor | Fixed-point scale | Current-layer post-processing |
| Mul_prev | ONNX quantization-related tensor | Fixed point by input channel and fused | Internal compensation in the W1A8 PE |
| Conv11 weights and biases | ONNX initializer | Q1.15 / Q4.12 | Final 1×1 detection head |
| COE files | Parameter extraction script output | Written in ROM address order | BRAM ROM initialization |

# 5 FPGA Hardware Architecture and Datapath Design

## 5.1 Overall Data Flow

The top module Yolo3-tiny_Top receives an RGB stream through a valid/ready handshake and outputs 32-bit serial detection results. The design follows a streaming style, and ready signals provide backpressure between modules, reducing dependence on external storage.

The image passes through the RGB adapter, Padding Adapter, and LineBuffer_3x3 to form 3×3 windows. Standard or W1A8 PEs process the windows, followed by post-processing, clipping, and optional 2×2 MaxPool. Conv11 uses PE_NUM=5 and serializes the final 10×10×75 tensor in y/x/channel order.

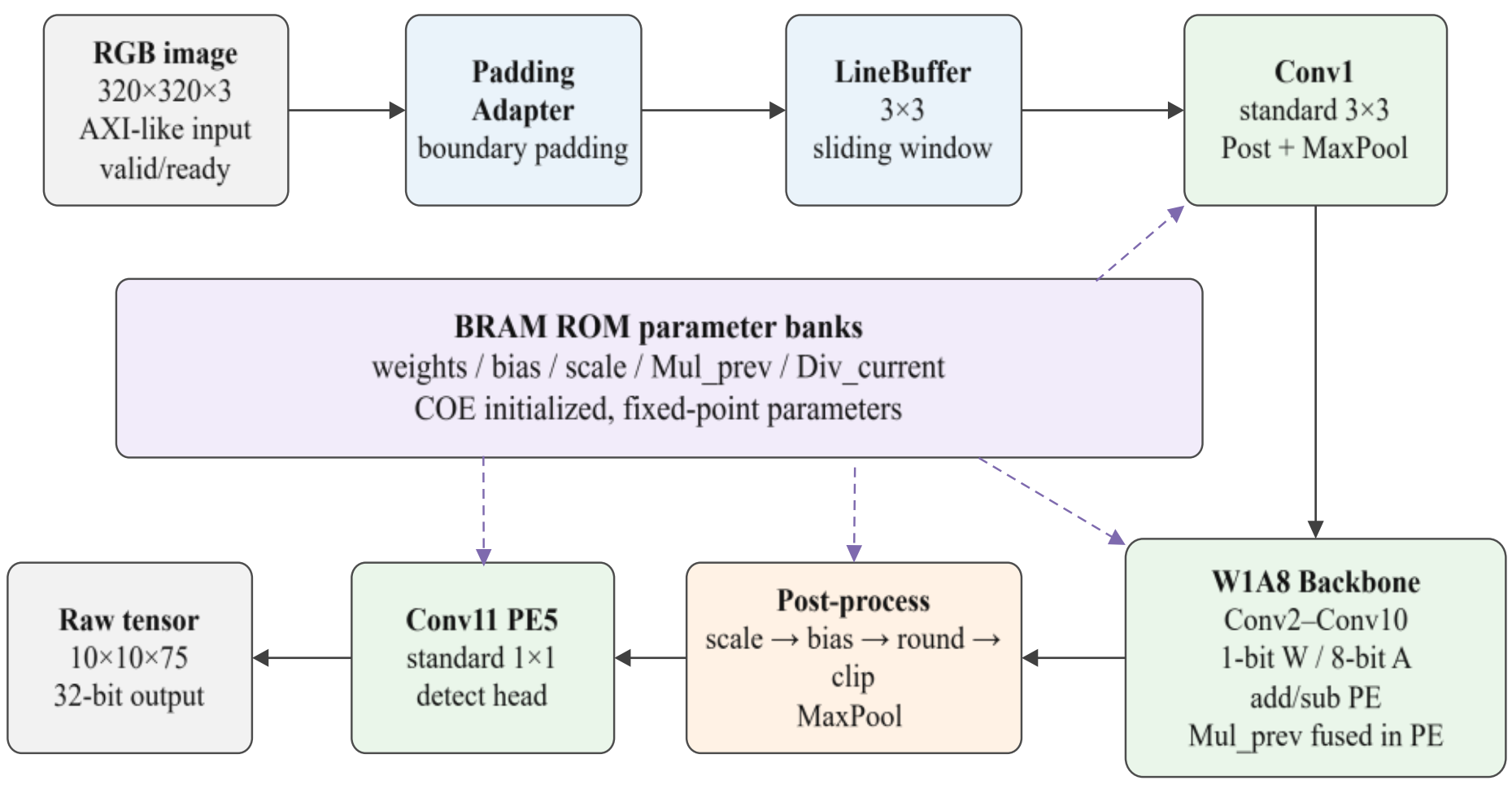

Fig. 1. RTL data-flow diagram of the W1A8 YOLOv3-tiny-like detector on PYNQ-Z2

## 5.2 Core Modules

Table 4. Main RTL modules and implementation points

| Module | Function | Implementation point |
|---|---|---|
| RGB input interface | Receives a 320×320 RGB image stream | valid/ready handshake |
| Padding Adapter | Generates convolution boundary padding | Supports 3×3 convolution with unchanged spatial size |
| LineBuffer_3x3 | Forms a 3×3 sliding window | Reuses line buffers and reduces external memory access |
| W1A8 convolution PE | Performs binary-weight convolution accumulation | Replaces multiplication with sign-controlled add/sub accumulation and fuses Mul_prev |
| Standard convolution PE | Performs Conv1 and Conv11 standard convolution | Supports specified Q-format weights and biases |
| Post-process | Performs scaling, bias correction, and clipping | Aligns with ONNX quantized activations |
| MaxPool | Performs 2×2 downsampling | Enabled only after specified layers |
| BRAM ROM | Stores COE-initialized parameters | State machine is designed according to ROM read latency |
| Conv11 PE | Computes the final detection head | Computes 5 output channels in parallel each time |

The Padding Adapter and LineBuffer_3x3 generate padded 3×3 sliding windows in a streaming manner. This avoids repeated full-feature-map access and reduces external-memory pressure.

The W1A8 convolution PE accumulates or subtracts 8-bit activations according to the binary weight sign. When Mul_prev is indexed by input channel, the PE also performs channel-wise compensation during accumulation.

The post-processing module performs fixed-point scaling, bias correction, rounding, and clipping to align activations with ONNX quantization nodes. MaxPool is enabled only after specified layers. Parameters are initialized into BRAM ROMs through COE files, and the state machine follows ROM read latency.

The final Conv11 detection head computes 75 output channels with PE_NUM=5. It processes channels in groups and serializes 10×10×75 raw detection values in y/x/channel order.

# 6 Experiments and Results

## 6.1 Model Accuracy and Complexity

Table 5 compares the proposed W1A8 quantized YOLO model with representative lightweight detectors on the PASCAL VOC 2007 test set. The metrics include mAP50, GFLOPs, parameter count, and input size. Because the models use different input sizes and structures, the table mainly reflects their accuracy-complexity trade-off.

The proposed model uses 0.098 GFLOPs and 0.74 M parameters. Compared with YOLOv5n, it uses about 1/45.9 of the computation and 1/2.6 of the parameters. Its

mAP50 is 39.6%, which is lower than most recent lightweight detectors in Table 5, but still higher than the early YOLOv2-tiny model by 4.2 percentage points. This result indicates that the proposed model sacrifices part of the detection accuracy for extremely low computational and storage costs, while still retaining a basic detection capability under severe FPGA resource constraints.

Figure 2 compares Params-mAP50 and FLOPs-mAP50. The proposed model lies in the lower-left region of both planes, indicating very low storage and computation while preserving basic detection capability. This makes it suitable for memory- and compute-limited FPGA platforms.

Table 5. Accuracy and complexity comparison of lightweight object detection models

| Model | mAP50 (%) ↑ | GFLOPs ↓ | Params (M) ↓ | Input size |
|---|---|---|---|---|
| MobileNet-SSD[23] | 68.0 | 2.3 | 5.8 | 300×300 |
| YOLOv5s[24] | 74.4 | 16.5 | 7.2 | 640×640 |
| YOLOv5n[25] | 67.8 | 4.5 | 1.9 | 640×640 |
| YOLOv3-tiny[26] | 50.8 | 5.6 | 8.7 | 416×416 |
| YOLOv4-tiny[27] | 59.2 | 6.9 | 6.0 | 416×416 |
| YOLOv2-tiny[4] | 35.4 | 7.0 | 15.86 | 416×416 |
| Ours | 39.6 | 0.098 | 0.74 | 320×320 |

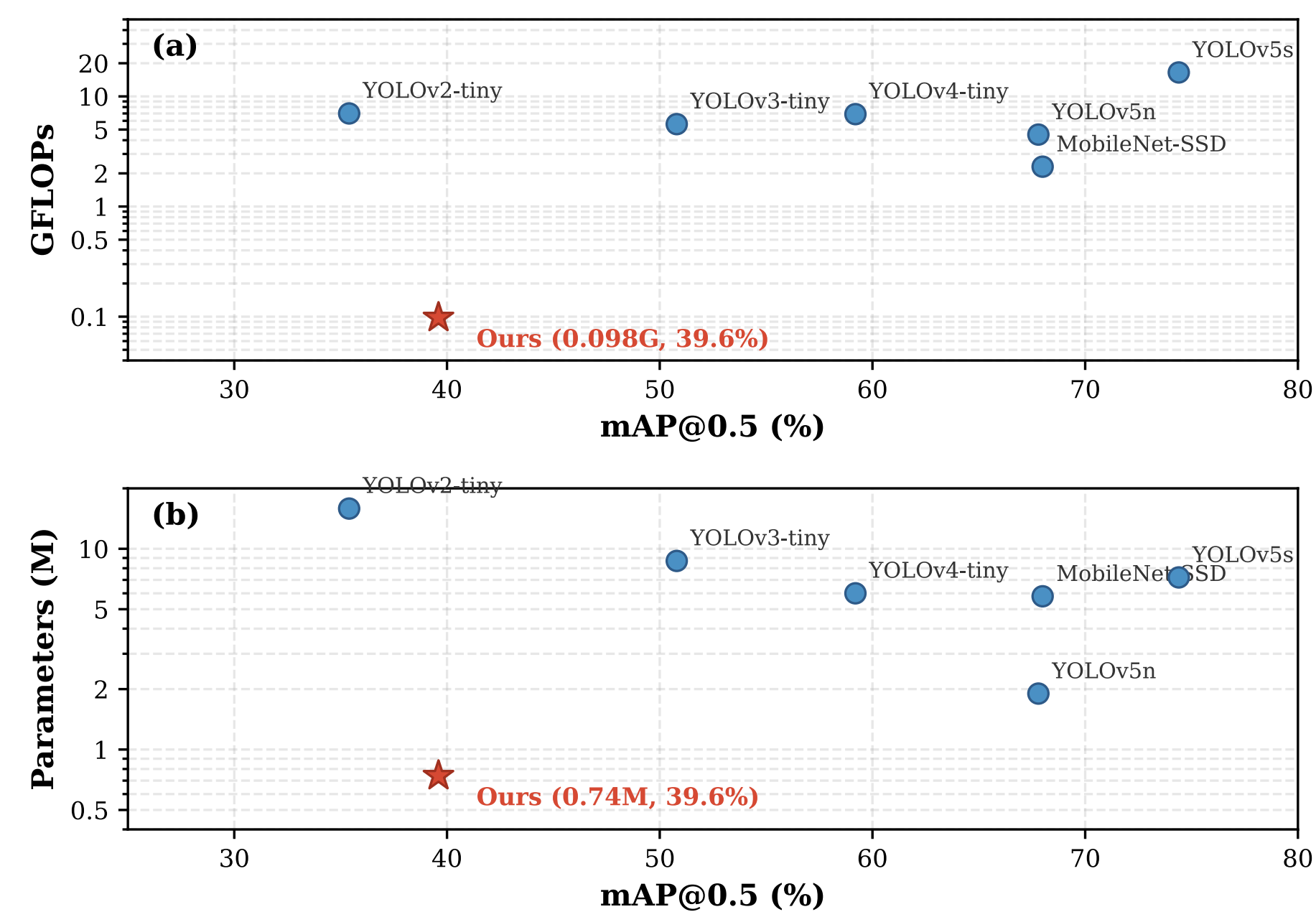

Fig. 2. Accuracy-complexity comparison of lightweight object detection models: (a) GFLOPs versus mAP50; (b) parameters versus mAP50.

### 6.2 Detection Results

Figure 3 visualizes detection results from the FPGA RTL data path after decoding, confidence filtering, and NMS. With 0.74 M parameters, the model detects medium and large objects such as persons, vehicles, and animals. The single 10×10 detection head limits small-object localization and recall, and binary weights with 8-bit activations can lead to low-confidence boxes or occasional false detections.

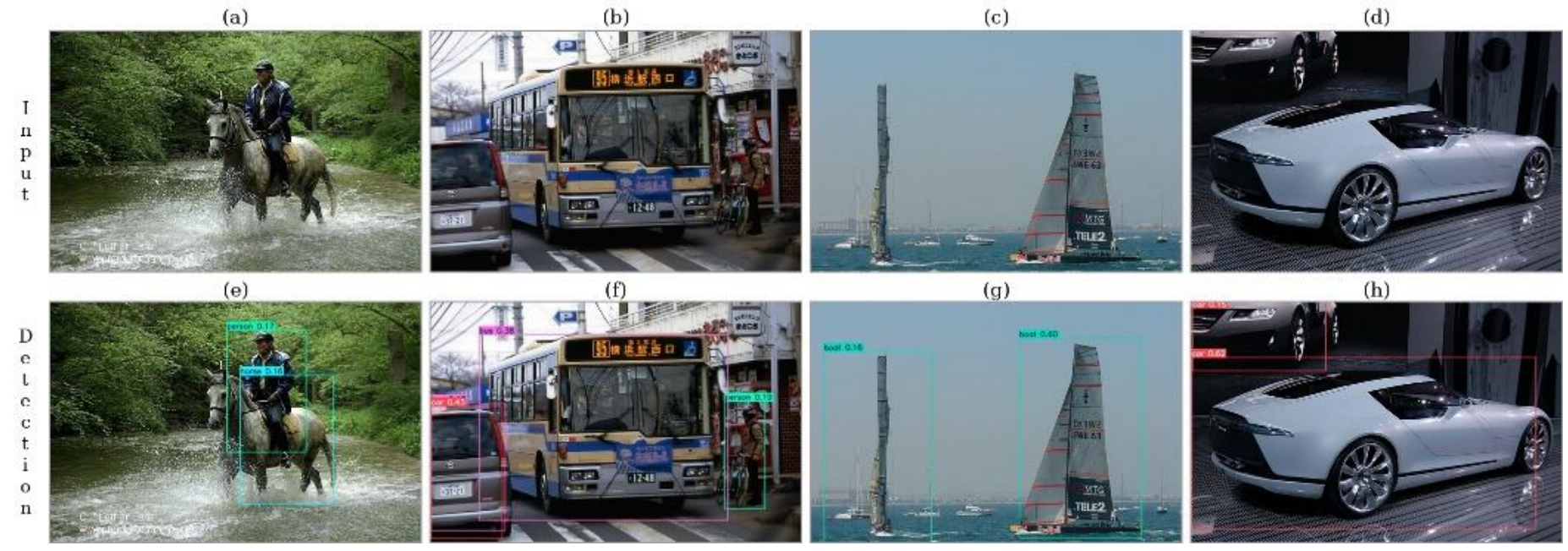

Fig. 3. Visualization of W1A8 quantized detection results based on FPGA RTL output

### 6.3 Numerical Consistency Verification Between RTL and ONNX

To verify RTL computation and parameter packing, a Vivado simulation testbench is built with a 320×320 RGB input sample. ONNX Runtime provides the software reference. RTL outputs are converted to floating point according to their Q formats and compared with ONNX intermediate nodes layer by layer.

Table 6. Numerical consistency comparison between RTL outputs and ONNX intermediate nodes

| Verification target | Software reference position | Data format | max abs | mean abs | corr. |
|---|---|---|---|---|---|
| Conv1 raw | Conv1 pre-activation output | DUT/2^19 | 0.132158 | 0.001345 | corr=0.999999 |
| Conv1 post | Conv1 post-quantization output | 8-bit quantized value | 4 | 0.121686 | 98.8094% of outputs within 1 LSB |
| Conv2 post | Conv2 post-quantization output | 8-bit quantized value | 6 | 0.231006 | 98.7648% of outputs within 1 LSB |
| Final raw conv | Conv11 detection-head output | 32-bit, 15 fractional bits | 0.10889649 | 0.020027 | corr=0.999964 |

Table 6 reports Conv1 pre-activation, Conv1 post-quantization, Conv2 post-quantization, and Conv11 detection-head outputs. The RTL results are highly consistent with ONNX. The final raw output reaches a correlation coefficient of 0.999964 and a mean absolute error of 0.020027, confirming the correctness of parameter extraction, fixed-point conversion, ROM access, and RTL computation.

### 6.4 FPGA Resources and Performance

The design is synthesized and implemented in Vivado to evaluate resource use on PYNQ-Z2. The reported metrics include LUT, LUTRAM, FF, BRAM, DSP, IO, BUFG, MMCM, power, and performance. Table 7 gives the implementation results.

Table 7. FPGA resource utilization and power estimate

| Category | Metric | Value | Description |
|---|---|---|---|
| Resource | LUT | 61% | Mainly used for control logic, add/sub accumulation, and data paths |
| Resource | LUTRAM | 32% | Used for small buffers or distributed storage |
| Resource | FF | 14% | Used for pipeline registers and state registers |
| Resource | BRAM | 73% | Mainly used for parameter ROMs, line buffers, and intermediate buffers |
| Resource | DSP | 91% | Mainly used for standard convolution, the detection head, and fixed-point scaling |
| Resource | IO | 13% | Input and output interface usage |
| Resource | BUFG | 16% | Global clock resource usage |
| Resource | MMCM | 25% | Clock-management resource usage |
| Power | Total On-Chip Power | 2.047 W | Vivado power estimate |
| Power | Dynamic Power | 1.881 W | 92% of total on-chip power |
| Power | Static Power | 0.166 W | 8% of total on-chip power |
| Temperature | Junction Temperature | 48.6 ℃ | Vivado estimated junction temperature |
| Performance | Maximum clock frequency | 30 MHz | Determined by post-implementation timing |
| Performance | Cycles per frame | 23.79 M | Measured from runtime statistics |
| Performance | PL inference latency | 793.17 ms | Converted from cycle count and 30 MHz clock |
| Performance | PS processing latency | 2.41 ms | Includes input-output and post-processing overhead |
| Performance | Latency per frame | 795.58 ms | Sum of PL inference latency and PS processing latency |
| Performance | FPS | 1.26 frames/s | Converted from latency per frame |

Table 7 shows that BRAM and DSP are the tightest resources. BRAM utilization is 73%, mainly from parameter ROMs, line buffers, and intermediate buffers. DSP utilization is 91%, mainly from Conv1, Conv11, and fixed-point scaling. W1A8 reduces DSP pressure in the main body, but fixed-point standard convolution and post-processing still consume many DSPs.

The maximum frequency is 30 MHz, and one frame requires 23.79 M cycles. The PL inference latency is 793.17 ms, the PS processing latency is 2.41 ms, and the end-to-end latency is 795.58 ms, corresponding to 1.26 frames/s. This result demonstrates the feasibility of the full RTL deployment and end-to-end execution on the target FPGA. Nevertheless, the current throughput remains limited by the 30 MHz timing closure, PE parallelism, the detection-head critical path, and inter-layer scheduling. These bottlenecks will be addressed in future optimization passes.

# 7 Conclusion and Future Work

This paper implements an RTL data path for a W1A8 quantized YOLOv3-tiny-like detector on a low-cost FPGA. From a trained ONNX model, the design completes parameter extraction, fixed-point conversion, channel-wise scale fusion, COE generation, and BRAM ROM deployment. The Verilog RTL includes input adaptation, padding, line buffering, W1A8 convolution, post-processing, max pooling, and the final detection head.

Experiments show 39.6% mAP50 on VOC with 0.098 GFLOPs and 0.74 M parameters. RTL verification gives a correlation coefficient of 0.999964 and a mean absolute error of 0.020027 for the final raw output against ONNX. The implementation closes timing on XC7Z020 with BRAM and DSP utilization at 73% and 91% respectively; frequency and throughput remain the principal optimization targets for future work.

Future work will improve both the model and the hardware. A lighter multi-scale detection head or feature-fusion structure can improve small-object recall. Higher PE parallelism, better fixed-point scaling, improved BRAM allocation, and more efficient inter-layer scheduling can further reduce resource pressure and improve throughput.